# Brain-inspired computing – We need a master plan.

*A. Mehonic & A.J. Kenyon*
Department of Electronic & Electrical Engineering
UCL
Torrington Place
London
WC1E 7JE
United Kingdom

## Abstract

New computing technologies inspired by the brain promise fundamentally different ways to process information with extreme energy efficiency and the ability to handle the avalanche of unstructured and noisy data that we are generating at an ever-increasing rate. To realise this promise requires a brave and coordinated plan to bring together disparate research communities and to provide them with the funding, focus and support needed. We have done this in the past with digital technologies; we are in the process of doing it with quantum technologies; can we now do it for brain-inspired computing?

## Main

### The problem

Our modern computing systems consume far too much energy. They are not sustainable platforms for the complex Artificial Intelligence (AI) applications that are increasingly a part of our everyday lives. We usually don't see this, particularly in the case of cloud-based systems, as we are usually more interested in their functionality – how *fast* are they; how *accurate*; how many parallel operations per second? We are so accustomed to having information available near-instantaneously that we don't think about the energy, and therefore environmental, consequences of the computing systems giving us this access. Nevertheless, each Google search has a cost: data centres currently use around 200 terawatt hours of energy per year, forecast to grow by around an order of magnitude by the end of the decade[1].

It is true that not all data-intensive computing requires Machine Learning (ML) or AI, but we are seeing AI deployed so widely that we must be concerned about its environmental cost. We should also consider applications such as the Internet of Things (IoT) and autonomous robotic agents that may not need always to be AI-enabled but must still reduce their energy consumption. The vision of the IoT cannot be achieved if the energy requirements of the myriad connected devices are too high. Recent analysis shows that increasing demand for computing power vastly outpaces improvements made through Moore's law scaling[2]. Computing power demands now double every two months (Figure 1a). Nevertheless, remarkable improvements have been made through a combination of smart architecture and software-hardware co-design. For example, the performance of NVIDIA GPUs has improved by the factor of 317 since 2012: far beyond what would be expected from Moore's law alone (Figure 1b). Further impressive performance improvements have been demonstrated at the R&D stage and it is likely that we can achieve more. Unfortunately, it is unlikely that digital solutions alone will cope with demand over an extended period. This is especially apparent when we consider the shockingly high cost of training required for the most complex ML models (Figure 1c). Alternative approaches are needed.

Fundamentally, the energy problem is a consequence of digital computing systems storing data separately from where they are processed. This is the classical von Neumann architecture, underpinning digital computing systems, in which processors spend most of their time and energy moving data. Luckily, we can improve the situation by taking inspiration from biology, which takes a different approach entirely. There is a system that achieves both energy efficiency and advanced functionality remarkably well: the human brain. Recognising that we still have much to learn about how the brain operates and that our aim is not simply to emulate biological systems, we can nevertheless learn from the significant progress achieved in neuroscience and computational neuroscience in the last few decades. We know just enough about the brain to use it as an inspiration to drive technology improvements.

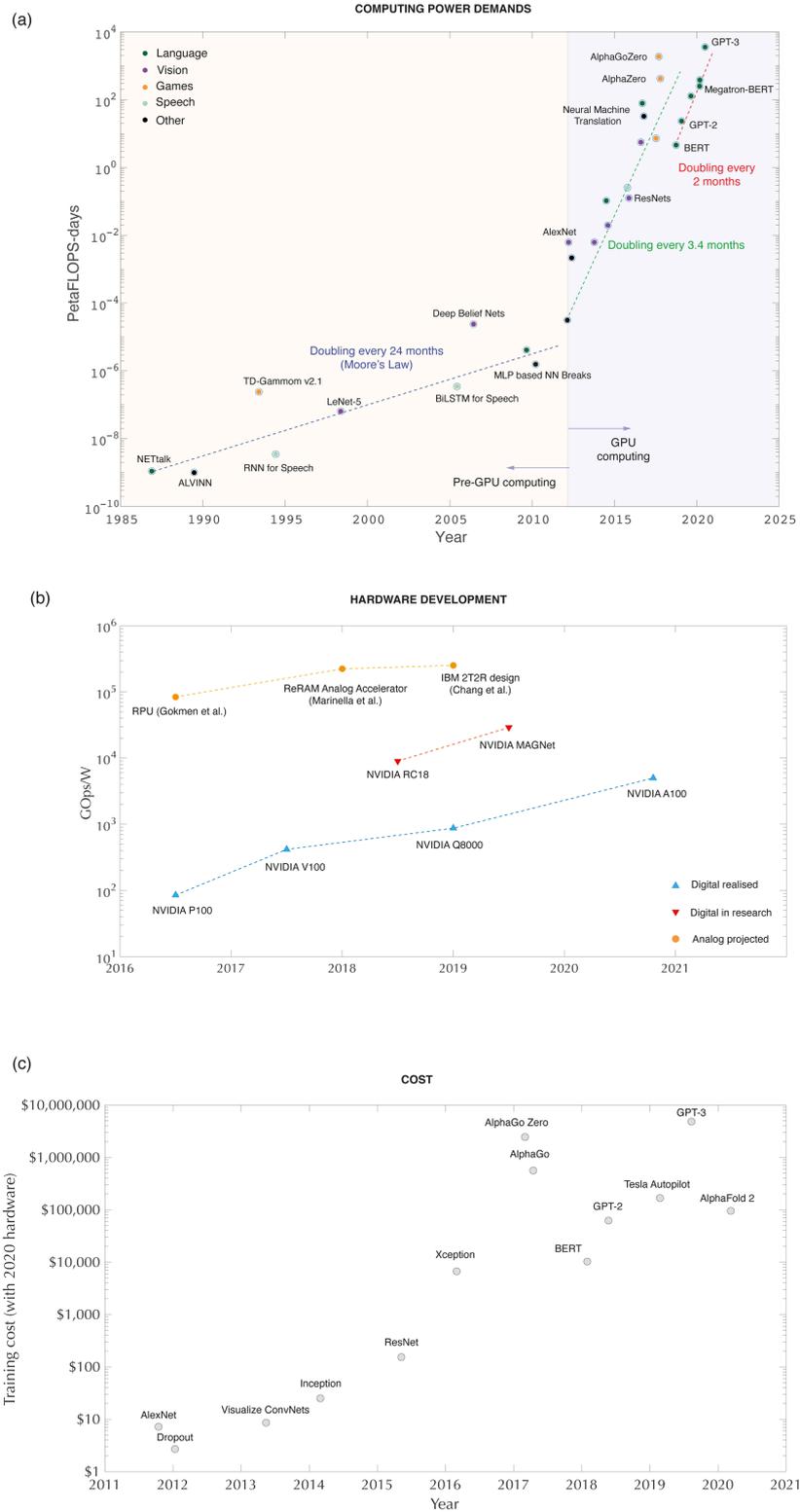

*Figure 1. (a)* The increase in computing power demands over the past four decades expressed in PetaFLOPS-days. Until 2012, computing power demand doubled every 24 months; recently this has shortened to approximately every two months. The colour legend indicates different application domains. Data obtained from [2]. *(b)* Improvements in AI hardware efficiency over the last five years. State-of-the-art solutions have driven increases in computing efficiency of over 300 times. Solutions in research and development promise further improvements. *(c)* Increase since 2011 of the costs of training AI models. Such an exponential increase is clearly unsustainable. Data obtained from [3].

> **BOX 1 – What do we mean by "neuromorphic" systems?**
>
> Taking inspiration from the brain allows us to approach information processing fundamentally differently to the way existing digital systems work. Different brain-inspired ("neuromorphic") platforms use different approaches: analogue data processing, asynchronous communication, massively parallel information processing or spiking-based information representation. These properties distinguish them from digital von Neumann computers.
>
> The term *neuromorphic* encompasses at least three communities of researchers, distinguished by whether their aim is to emulate neural function, simulate neural networks, or develop new electronic devices beyond traditional digital paradigms.
>
> *Neuromorphic engineering* takes inspiration from the way the brain uses the physics of biological synapses and neurons to "compute". Neuromorphic engineers work to emulate the functions of biological neurons and synapses. They harness the physics of analogue electronics – such as carrier tunnelling, charge retention on silicon floating gates, and the exponential dependence of various device or material properties on field – to define elementary operations to underpin audio or video processing or smart sensors, for example. Transistors are used as analogue circuit elements with rich dynamic behaviour rather than essentially binary switches.
>
> *Neuromorphic computing* looks to biology to inspire new ways to process data. Research is more application-driven and looks to simulate the structure and/or operation of biological neural networks. This may mean co-locating storage and computing, as the brain does, to avoid shuffling data constantly between processor and memory; or it may mean adopting wholly different ways of computing based on voltage spikes modelling the action potentials of biological systems.
>
> Underpinning everything are the devices and materials needed to implement bio-inspired functions. These are traditionally CMOS transistors, but recent developments promise new electronic and photonic devices whose properties we can tailor to mimic biological elements such as synapses and neurons. These *neuromorphic devices* could provide exciting new technologies to expand the capabilities of neuromorphic engineering and computing.
>
> Foremost amongst these new devices are *memristors*: two terminal electronic devices whose resistance is a function of their history. Their complex dynamic response to electrical stimulation means they can be used as digital memory elements, as variable weights in artificial synapses, as cognitive processing elements, optical sensors, and devices that mimic biological neurons[4]. They may embody some of the functionality of biological dendrites[5] and their dynamic response can generate oscillatory behaviour similar to that of the brain – controversially, operating on the edge of chaos[6,7]. They may also be linked with biological neurons in a single system[8]. Crucially, they do all of this while expending very little energy.

### Biological inspiration

Biology does not separate data storage from processing. The same elements – principally neurons and synapses – perform both functions in massively parallel and adaptable structures. The $10^{10}$ neurons and $10^{14}$ synapses contained in the typical human brain expend approximately 20 W of power, while a digital simulation of an artificial neural network of approximately the same size consumes 7.9 MW[9]. That six order of magnitude gap shows how much we must improve. The brain also processes with extreme efficiency data that are noisy, imprecise and unstructured; a task very costly in energy and time for even the most powerful digital supercomputers. There is a lesson to learn: brain-inspired, or *neuromorphic*, computing systems could transform the way we process data, both in terms of energy efficiency and of their capacity to handle real-world uncertainty.

This is not a new idea. The term *neuromorphic*, describing devices and systems that mimic some functions of biological neural systems, was coined in the late 1980s by Carver Mead at the California Institute of Technology[10,11]. The concept came from work undertaken over previous decades to model the nervous system as equivalent electrical circuits[12] and to build *analogue* electronic devices and systems to provide similar functionality (Box 1).

Of course, there is a good reason why we use a digital data representation for most applications: we want high precision, reliability and determinacy. Simple calculations should never give wrong answers. However, digital abstraction discards massive amounts of information, found in the physics of transistors, for the minimum information quantum: a single bit. And we pay a significant energy cost by trading efficiency for reliability. As AI applications are often probabilistic at heart we must consider if this trade-off makes sense and whether analogue and mixed digital-analogue systems would be better suited. The computational tasks underpinning AI applications are very compute-intensive (and therefore energy-hungry) when performed by digital computers with von Neumann architectures. However, we might perform similar tasks much more energy-efficiently on analogue or mixed systems that use a spiking-based representation of information, either solely for communication or to both compute and transfer data. There has therefore been a recent resurgence in interest in neuromorphic computing, driven by the growth in AI systems and by the emergence of new devices that offer new and exciting ways to mimic some of the capabilities of biological neural systems (Box 1).

Definitions of what makes a technology neuromorphic vary from discipline to discipline. Loosely speaking, the neuromorphic story is a hardware one: neuromorphic chips aim to integrate and utilise various useful features of the human brain, including in-memory computing, spike-based information processing, fine-grained parallelism, reduced precision and increased stochasticity, adaptability, learning in hardware, asynchronous communication, and analogue processing. While it is debatable how many of these need to be implemented for something to be classified as neuromorphic, this is clearly a different approach from digital AI. Nevertheless, we should not be lost in terminology; the main question is whether this approach is *useful*.

Approaches to neuromorphic technologies lie on a spectrum between reverse-engineering the structure and function of the brain (analysis), and living with our lack of knowledge of the brain but taking inspiration from what we do know (synthesis). Perhaps foremost among the former approaches is the Human Brain Project, a high-profile and hugely ambitious ten-year programme funded by the European Union from 2013. The programme has generated two hardware platforms – SpiNNaker (at Manchester) and BrainScaleS (at Heidelberg) – which implement highly complex silicon models of brain architectures to understand better the operation of the biological brain. At the other end of the spectrum numerous groups are augmenting the performance of digital or analogue electronics using selected biologically-inspired methods. Figure 2 summarises the range of existing neuromorphic chips, divided into four categories depending on their position on the analysis-synthesis spectrum and their technology platform. In all of this it is important to keep in mind that neuromorphic engineering isn't just about high-level cognitive systems, but also offering energy, speed and security gains in small-scale edge devices with limited cognitive abilities. Wherever along that scale we choose to be, taking some form of inspiration from biology has enormous promise.

## Prospects

The above is not to say that neuromorphic systems will, or should, replace digital computation. Instead, precision calculations, accurately performed by existing systems, should remain the preserve of digital computation while neuromorphic systems can process unstructured data, perform image recognition, classification of noisy and uncertain data sets, and underpin novel learning and inference systems. In autonomous and IoT-connected systems, they can provide huge energy savings over their digital counterparts. We must also include quantum computing in this vision. A practical quantum computer, while still several years away by any estimation, would certainly revolutionise many computing tasks. However, it is unlikely that IoT-connected smart sensors, edge computing devices or autonomous robotic systems will adopt quantum computing without depending on cloud computing. There will remain a need for low-power computing elements capable of dealing with uncertain and noisy data. We can imagine

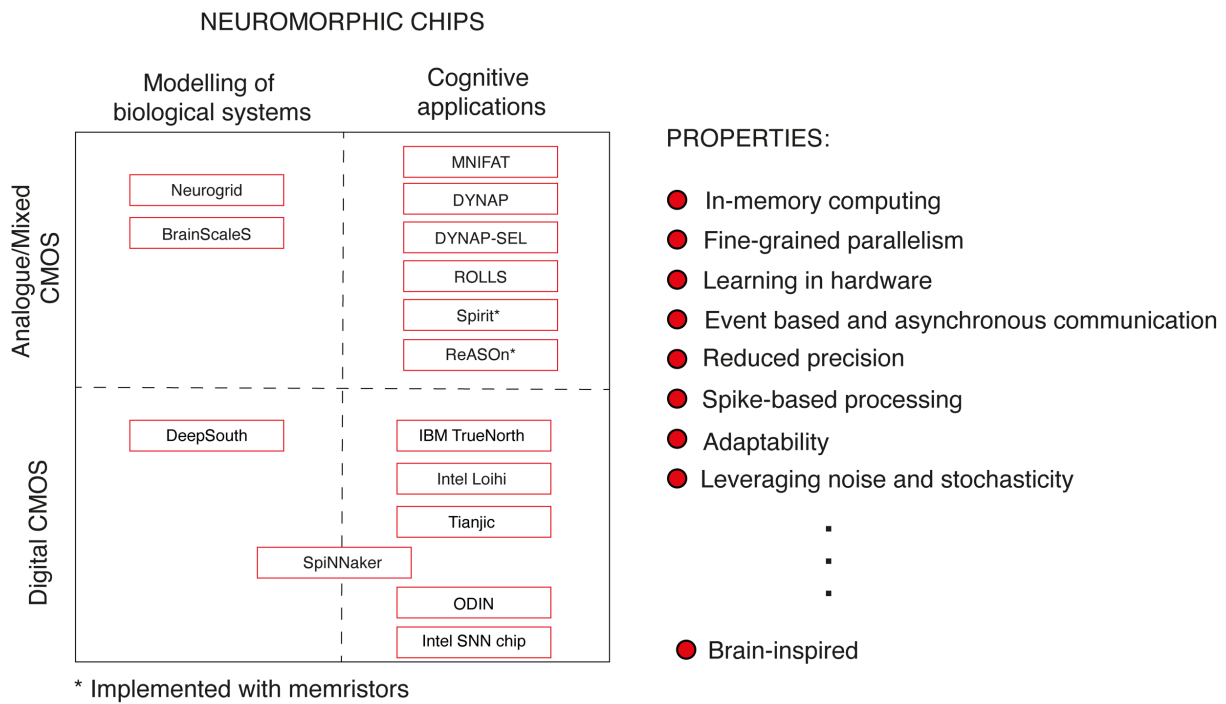

*Figure 2:* Neuromorphic chips can be classified as either modelling biological systems or applying brain-inspired principles to novel computing applications. They may be further subdivided into those based on digital CMOS with novel architecture (for example, spikes may be simulated in the digital domain rather than implemented as analogue voltages) and those implemented using some degree of analogue circuitry. In all cases, however, they share at least some of the properties listed on the right-hand side, which distinguish them from conventional CMOS chips. Here we classify examples of recently developed neuromorphic chips. Further details of each can be found in the relevant reference: Neurogrid[13], BrainSclaseS[14], MNIFAT[15], DYNAP[16], DYNAP-SEL[17], ROLLS[18], Spirit[19], ReASOn[20], DeepSouth[21], SpiNNaker[22], IBM TrueNorth[23], Intel Loihi[24], Tianjic[25], ODIN[26], and the Intel SNN chip[27].

a three-way synergy between digital, neuromorphic and quantum systems (Figure 3).

Just as the development of semiconductor microelectronics relied on many different disciplines coming together, including solid state physics, electronic engineering, computer science, and materials science, neuromorphic computing is profoundly cross- and inter-disciplinary. Physicists, chemists, engineers, computer scientists, biologists, neuroscientists, all play key roles. Simply getting researchers from such a diverse set of disciplines to speak a common language is challenging. In our own work we spend considerable time and effort ensuring that everyone in the room understands terminology and concepts in the same way. A case for bridging the communities of computer science (specifically artificial intelligence) and neuroscience (initially computational neuroscience) is clear. After all, many concepts found in today's state-of-the-art AI systems arose in the 1970s and 80s in neuroscience communities. Of course, AI systems do not need to be completely bio-realistic or capture everything from neuroscience. We must go further and include other disciplines in these discussions, recognising that many of the strides we have made in AI or neuroscience have been enabled by different communities – e.g. innovations in material science, nanotechnology, or electrical engineering. Further, conventional CMOS technology may not be the best fabric to efficiently implement new brain-inspired algorithms; innovations across the board are needed.  Engaging these communities early means reduces the risk of wasting effort on directions that have already been explored and failed, or of reinventing the wheel.

## Seizing the opportunity

So, if neuromorphic computing is needed, how to achieve it? First, the technical requirements. Bringing together diverse research communities is necessary but not sufficient. Incentives, opportunities and infrastructure are needed. The neuromorphic community is a disparate one lacking the focus of the quantum computing community, or the clear roadmap of the semiconductor industry. Initiatives around the globe are starting to gather the required expertise, and early stage momentum is building. How can we build on this? Funding is key. Investment in neuromorphic research is nowhere near the scale of that in digital AI or quantum technologies (Box 2). In one sense that is not surprising given the maturity of digital semiconductor technology, but it is a missed opportunity. While there are a few examples of medium-scale investment in neuromorphic R&D such as the IBM AI Hardware Centre's range of brain-inspired projects (including the TrueNorth chip), Intel's development of the Loihi processor, and the US Brain Initiative project, the sums committed are well below what they should be given the promise of the technology to disrupt digital AI.

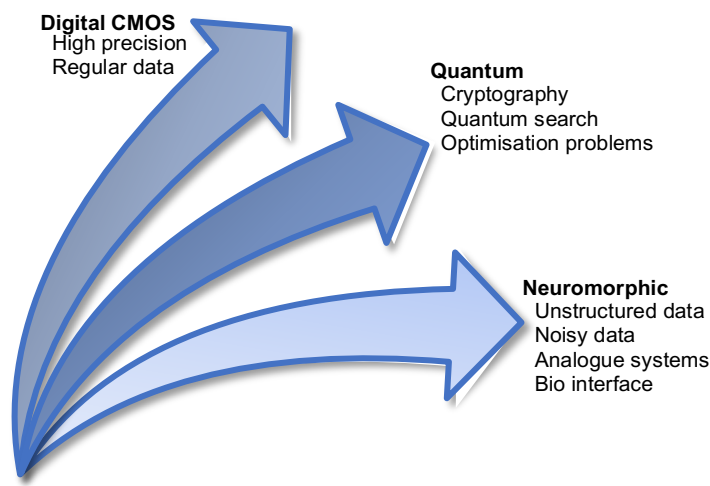

*Fig. 3:* A scenario for the future of computing hardware. Digital CMOS, Quantum and Neuromorphic systems can operate in parallel, each in different applications domains. Of the three, neuromorphic has received the least attention.

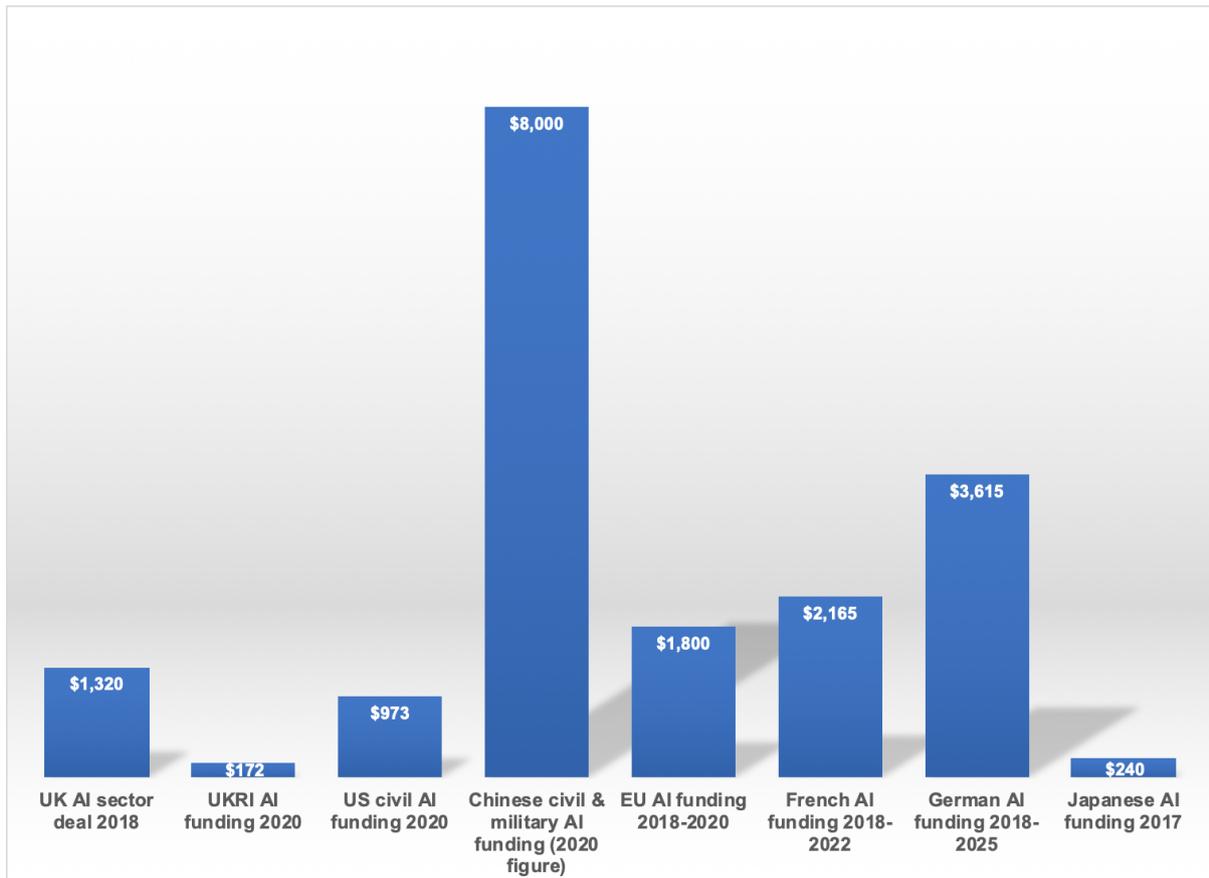

*Figure 4: A comparison of recent global public research funding of digital AI technologies. Figures are in US dollar equivalent (2021 exchange rate) and are expressed as millions of dollars. While some are in-year snapshots (eg UKRI funding committed for 2020) and others are for multi-year programmes, the figure illustrates the scale of public funding in digital technologies. Disruption of the AI ecosystem by the development of efficient neuromorphic technologies would put much of this investment at risk.*

The time is ripe for bold initiatives. At a national level, governments need to work with academic researchers and industry to create mission-oriented research centres to accelerate the development of neuromorphic technologies. This has worked well in areas such as quantum technologies and provides focus and stimulus. Such centres may be physical or virtual but must bring together the best researchers across diverse fields. Their approach must be different from that of conventional electronic technologies in which every level of abstraction (materials, devices, circuits, systems, algorithms and applications) is designed and optimised by a different community. We need holistic and concurrent design across the whole stack. It isn't enough for circuit designers to consult computational neuroscientists before designing systems; engineers and neuroscientists must work together throughout the process to ensure as full an integration of bio-inspired principles into hardware as possible. Interdisciplinary co-creation must be at the heart of our approach. Research centres must house a broad constituency of researchers.

Alongside the required physical and financial infrastructure, we need a trained workforce. Electronic engineers are rarely exposed to ideas from neuroscience, and vice-versa. Circuit designers and physicists may have a passing knowledge of neurons and synapses but are unlikely to be familiar with cutting edge computational neuroscience. There is a strong case to set up Masters courses and doctoral training programmes to develop neuromorphic engineers. UK research councils sponsor Centres for Doctoral Training (CDTs) – focused programmes supporting specific areas that have an identified need for trained researchers. CDTs can be single- or multi-institution; there are significant benefits to institutions collaborating on these programmes, in terms of student experience and in creating complementary teams across institutional boundaries. Programmes generally work closely with industry and build cohorts of highly skilled researchers in ways that more traditional doctoral programmes often do not. There is a good case to be made to develop CDTs, or something similar, to stimulate interaction between nascent neuromorphic engineering communities and provide the next generation of researchers and research leaders. Pioneering examples include the Groningen Cognitive Systems and Materials research programme, which aims to train tens of doctoral students specifically in materials for cognitive (AI) systems[28]; ETH Zurich work on analogue circuit design for neuromorphic engineering[29]; large-scale neural modelling at Stanford University[30], and development of visual neuromorphic systems at the Instituto de Microelectrónica de Sevilla[31]. There is scope to do much more.

> **BOX 2 – The AI funding landscape**
>
> Investment in "conventional" digital AI is booming, fuelled by the need to process ever-increasing volumes of data, and the development of hardware to support existing compute- and memory-intensive algorithms. The UK government announced in April 2018 a £950 million "sector deal" in digital AI, in addition to existing research council support. France announced a €1.8 billion government investment in AI from 2018 to 2022[32], Germany committed €3 billion from 2018 to 2025, while Japan spent ¥26 trillion in 2017. US government funding of civil AI technologies was $973 million in 2020[33]; figures are harder to come by for US military AI funding, as non-AI projects are often included in published analysis. China is estimated to be investing up to $8 billion in both civil and military AI and is constructing a $2.1 billion AI research park near Beijing[34], while the European Commission committed €1.5 billion in the period 2018-2020[35]. Commercial investment dwarfs this. In the USA one estimate puts the total investment in AI companies in 2019 at $19.5 billion[36], and global investment is predicted to be around $98 billion by 2023[37]. Such sums must be considered at risk if our current hardware systems cannot support potentially disruptive neuromorphic algorithms and architectures. If neuromorphic technologies offer anything like the efficiency savings and enhanced performance they promise, smart money will hedge its bets on novel technologies and architectures alongside digital systems.
>
> Comparable figures are not available for neuromorphic technologies, as they currently lack focus and government-level visibility. Research funding is therefore piecemeal and at project, rather than strategic, level. While there have been various estimates published – for example, that the global neuromorphic chip market will grow from $111 million in 2019 to $366 million in 2025[38], the safest conclusion to draw is that funding of neuromorphic systems lags way behind that of digital AI or of quantum (of which more below).

Similar approaches could work at the trans-national level. As in any research field, collaboration is most successful when it is the best working with the best, irrespective of borders. In such an interdisciplinary endeavour as neuromorphic computing this is critical, so international research networks and projects will undoubtedly have a role to play. An early example is the Chua Memristor Centre at the University of Dresden[39], which brings together many of the leading memristor researchers across materials, devices and algorithms. Again, much more can and must be done.

How to make this attractive to governments? Government commitment to more energy-efficient bio-inspired computing can be part of a broader large-scale decarbonisation push. This will not only address climate change but will also provide technologies to accelerate the emergence of new, low-carbon, industries around big data, IoT, healthcare analytics, modelling for drug and vaccine discovery, and robotics, amongst others. If existing approaches to these industries rely on ever more large-scale digital data analysis, they increase their energy cost while offering sub-optimal performance. We can instead create a virtuous circle in which we greatly reduce the carbon footprint of the knowledge technologies that will drive the next generation of disruptive industries and, in doing so, we seed a host of new neuromorphic industries.

If we think this a tall order, consider quantum technologies, a sector that governments have invested in heavily. In the UK the government has so far committed around £1 billion to a range of quantum initiatives, largely under the umbrella of the National Quantum Technologies Programme. A series of research hubs, bringing together industry and academia, translate quantum science into technologies targeted at sensors and metrology, imaging, communications, and computing. A separate National Quantum Computing Centre builds on the work of the hubs and other researchers to deliver demonstrator hardware, software, and technology horizon-scanning to develop a general purpose quantum computer.

China has established a multi-billion dollar Chinese National Laboratory for Quantum Information Sciences, while the USA in 2018 commissioned a National Strategic Overview for Quantum Information Science[40], on the back of which it announced a 5-year $1.2 billion investment in quantum, on top of supporting a range of national quantum research centres[41]. Thanks to this research work there has been a global rush to start up quantum technology companies. One analysis found that in 2017 and 2018 funding for private companies reached $450 million[42]. No such joined-up support exists for neuromorphic computing, despite the technology being more established than quantum, and despite its potential to disrupt existing AI technologies on a much shorter time horizon. Of the three strands of future computing in our vision, neuromorphic is woefully under-invested.

Our message about how to exploit the potential of neuromorphic systems is clear: Step up investment in collaborative research through the establishment of research centres of excellence; provide agile funding mechanisms to enable rapid progress; provide mechanisms for close collaboration with industry to bring in commercial funding and generate new spin-outs and start-ups, similar to schemes already in place for quantum tech; develop training programmes for the next generation of neuromorphic researchers and entrepreneurs; and do all of this quickly and at scale.

Finally, a few words about what bearing the COVID-19 pandemic might have on our arguments. There is a growing consensus that the crisis has accelerated many developments already under way: for example, the move to more homeworking. While reducing the need for commuting and travel has direct benefits – some estimates put the reduction in global $CO_2$ as a result of the crisis at up to 17%[43] – new ways of working have a cost. To what extent will carbon savings from reduced travel be offset by increased date centre emissions? Addressing these will contribute to the larger decarbonisation agenda. So, if anything, the COVID pandemic has further emphasised the need to develop low carbon computing technologies such as neuromorphic systems.

Neuromorphic computing has the potential to transform our approach to AI. Thanks to the conjunction of new devices and technologies alongside a massive and growing demand for efficient AI systems we have a timely opportunity. Bold thinking is needed, and bold initiatives to support this thinking. Will we seize the opportunity?

## Competing interests

The authors are founders and directors of Intrinsic Semiconductor Technologies Ltd (www.intrinsicst.com), a spin-out company commercialising silicon oxide RRAM.